\documentclass[pra,twocolumn,nofootinbib,superscriptaddress]{revtex4-1}
\usepackage{graphicx}
\usepackage{dcolumn}
\usepackage{bm,bbm}
\usepackage{xcolor}
\usepackage{tcolorbox}
\usepackage{algorithm}
\usepackage{algpseudocode}
\usepackage{amsmath}
\usepackage{bbold}
\usepackage{braket}
\usepackage{nameref}
\usepackage[toc,page]{appendix}

\usepackage[colorlinks,breaklinks,linkcolor={blue},citecolor={magenta},urlcolor={blue}]{hyperref}
\usepackage{enumerate}

\makeatletter
\newcommand\org@hypertarget{}
\let\org@hypertarget\hypertarget
\renewcommand\hypertarget[2]{%
  \Hy@raisedlink{\org@hypertarget{#1}{}}#2%
  }
\makeatother

\begin{document}

\title{
Manipulating the symmetry of transverse momentum entangled biphoton states}

\author{Xiaoqin Gao}
\affiliation{Department of physics, University of Ottawa, Advanced Research Complex, 25 Templeton Street, K1N 6N5, Ottawa, ON, Canada}

\author{Yingwen Zhang}
\email{Yingwen.Zhang@nrc-cnrc.gc.ca}
\affiliation{National Research Council of Canada, 100 Sussex Drive, K1A 0R6, Ottawa, ON, Canada}

\author{Alessio  D'Errico}
\affiliation{Department of physics, University of Ottawa, Advanced Research Complex, 25 Templeton Street, K1N 6N5, Ottawa, ON, Canada}

\author{Felix Hufnagel}
\affiliation{Department of physics, University of Ottawa, Advanced Research Complex, 25 Templeton Street, K1N 6N5, Ottawa, ON, Canada}

\author{Khabat Heshami}
\affiliation{National Research Council of Canada, 100 Sussex Drive, K1A 0R6, Ottawa, ON, Canada}
\affiliation{Department of physics, University of Ottawa, Advanced Research Complex, 25 Templeton Street, K1N 6N5, Ottawa, ON, Canada}

\author{Ebrahim Karimi}
\affiliation{Department of physics, University of Ottawa, Advanced Research Complex, 25 Templeton Street, K1N 6N5, Ottawa, ON, Canada}
\affiliation{National Research Council of Canada, 100 Sussex Drive, K1A 0R6, Ottawa, ON, Canada}


\begin{abstract}
Bell states are a fundamental resource in photonic quantum information processing. These states have been generated successfully in many photonic degrees of freedom. Their manipulation, however, in the momentum space remains challenging. Here, we present a scheme for engineering the symmetry of two-photon states entangled in the transverse momentum degree of freedom through the use of a spatially variable phase object. We demonstrate how a Hong-Ou-Mandel interferometer must be constructed to verify the symmetry in momentum entanglement via photon ``bunching/anti-bunching'' observation. We also show how this approach allows generating states that acquire an arbitrary phase under the exchange operation.
\end{abstract}

\date{\today}
\maketitle

\section*{Introduction}
Quantum entanglement, considered one of the most counterintuitive features of quantum mechanics~\cite{Einstein}, is now one of the most important resources for quantum information tasks. In quantum optics it has been used as a fundamental tool in quantum cryptography~\cite{Ekert}, quantum dense coding~\cite{Bennett1992}, quantum teleportation~\cite{Bennett1993}, and quantum computation \cite{Raussendorf2001}. A great number of experiments have investigated the production of photonic entangled states, which have played a critical role in many important applications in quantum information processing. Photon pair generation through Spontaneous Parametric Down Conversion (SPDC) has been used to demonstrate entanglement in polarization~\cite{PhysRevA.45.7729}, path~\cite{PhysRevLett.77.1917}, spatial modes (e.g., Hermite-Gauss modes~\cite{PhysRevLett.90.143601, PhysRevA.94.033855}, Laguerre-Gauss modes~\cite{PhysRevA.89.013829,zhang2016engineering}), energy-time~\cite{PhysRevLett.62.2205} and time-bin~\cite{PhysRevLett.82.2594} degrees of freedom, and some of them simultaneously~\cite{PhysRevA.54.R1, PhysRevLett.92.210403, PhysRevLett.95.260501, wang2015quantum}. The SPDC state can also provide a good approximation of a momentum-position Einstein-Podolsky-Rosen (EPR) state when looking at the transverse momentum decomposition~\cite{PhysRevLett.92.210403}.

Momentum entanglement, as a continuous degree of freedom, can allow, in principle, to reach the ultimate limits of high-dimensional entanglement~\cite{PhysRevLett.92.210403}. In addition, high-dimensional quantum systems can allow entanglement to have high complexity and can be exploited for various quantum information tasks~\cite{2014, wang2018multidimensional, erhard2020advances}.
The momentum entangled state that naturally arises from SPDC is symmetric~\cite{PhysRevLett.92.233601}, i.e., the same state is obtained under the exchange of idler and signal photon. 
However, it is more challenging to generate antisymmetric momentum entanglement. One possible approach that exploits the pump symmetry has been explored in Ref.~\cite{PhysRevLett.90.143601}. The competition between the symmetry of polarization entanglement and pump shaping was shown to affect the two-photon interference behavior.

\begin{figure*}[t]
\includegraphics [width= 1\linewidth]{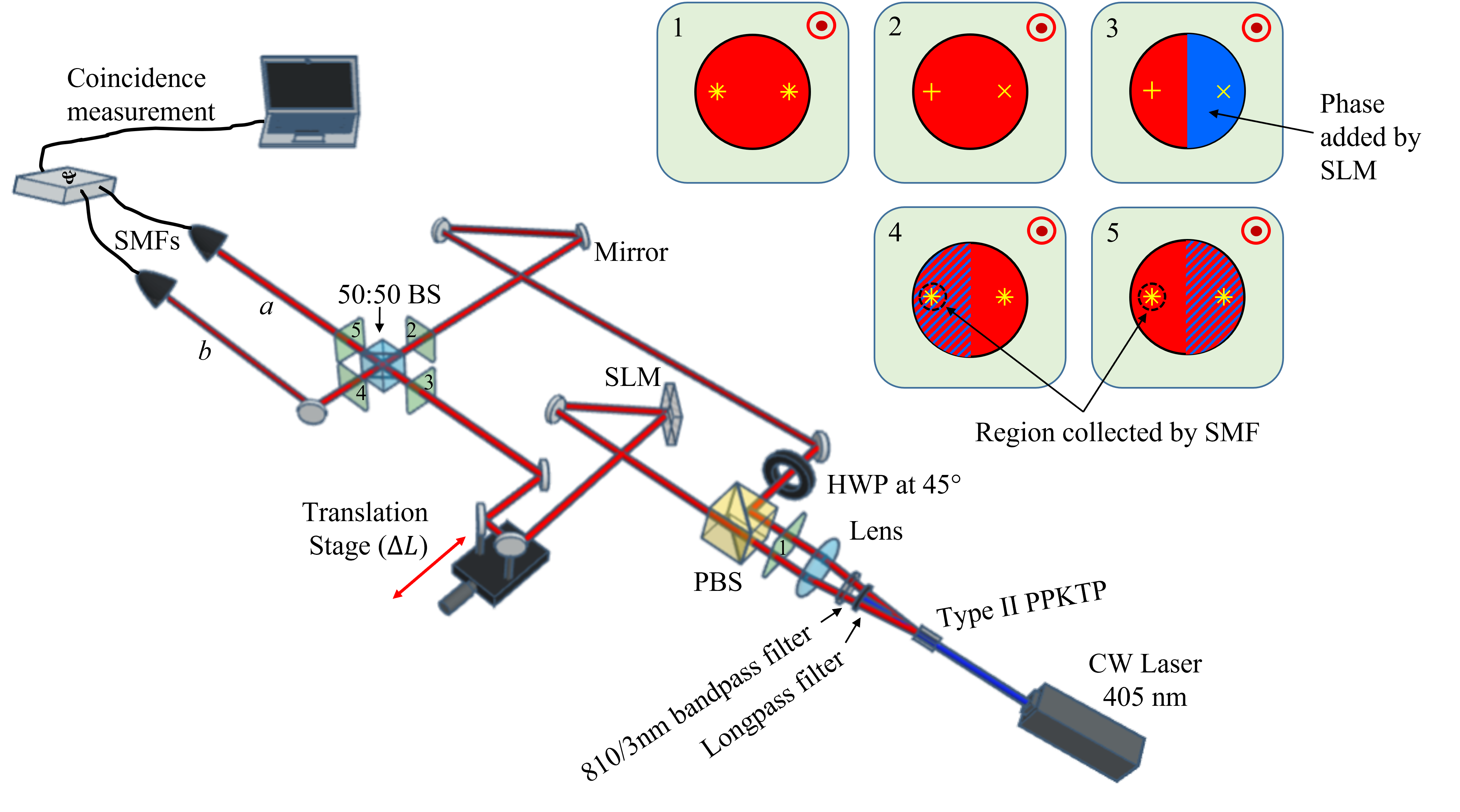}
\centering
\caption{{\bf Experimental setup for generating momentum entanglement by quantum interference.} A horizontally polarized continuous-wave pump beam (405\,nm, waist diameter 3.8\,mm) induces polarization-based SPDC in a Type-II periodically poled potassium titanyl phosphate (ppKTP) crystal ($5$ mm thick). A lens is used to map the {collinear} SPDC state in the transverse momentum degree of freedom. Correlated photons with orthogonal polarizations are then separated into different optical paths with a polarizing beamsplitter (PBS) after a longpass filter and 3\,nm bandpass filter. A half-wave plate (HWP) oriented at $45^{\circ}$ rotates the vertically-linear polarization (V) of the signal photon to horizontal polarization (H). The two photons are then made to interfere at a 50:50 beamsplitter (BS). A delay line on the idler path allows for adjustments of the optical path difference ${\Delta}L$ between the two photons. A spatial light modulator (SLM) is placed in the idler photon path before the BS to allow manipulation of the phase between the momentum entangled state.
The number of mirror reflections at the BS exit must have the same parity to have the two photons maintain momentum anti-correlated. Photons from opposite sides of the SPDC beam are collected by two single-mode fibers (SMFs) connected to avalanche photodiodes. Hong-Ou-Mandel interference can be observed after a coincidence measurement. The insets show the beam projections (beam coming out of the page) at the plane before the PBS (1) and the four numbered planes surrounding the 50:50 BS. (2,3) 
Before the BS with a phase applied in half of the beam (blue color) in (3). (4,5) 
After the BS showing how the two beams are overlapped and the region collected by the SMFs. Locations that are momentum correlated are marked with the same symbol ($+$ and $\times$).
}
\label{fig:setup}
\end{figure*}

Here, we demonstrate an approach to freely manipulate the relative phase defining the two-photon entangled state by introducing a spatially dependent phase distribution on one of the photons in the pair (namely, the idler) path. A phase jump applied between opposite values of the idler photon's transverse momentum affects the exchange symmetry. A $\pi$-phase jump allows conversion of the symmetric SPDC state into an antisymmetric one. Intermediate phase jumps will generate antisymmetric states, which gain a general phase factor under exchange operation. 
We also demonstrate how a two-photon interference setup needs to be constructed in order to verify the symmetry of such a momentum entangled state. This work provides a new method for quantum state engineering and entanglement verification in the momentum degree of freedom, which may be exploited in quantum imaging protocols, high-dimensional quantum communications, quantum information processing, and quantum simulations.

\section*{Scheme}
\begin{figure*}[tbph]
\includegraphics [width= 0.9\textwidth]{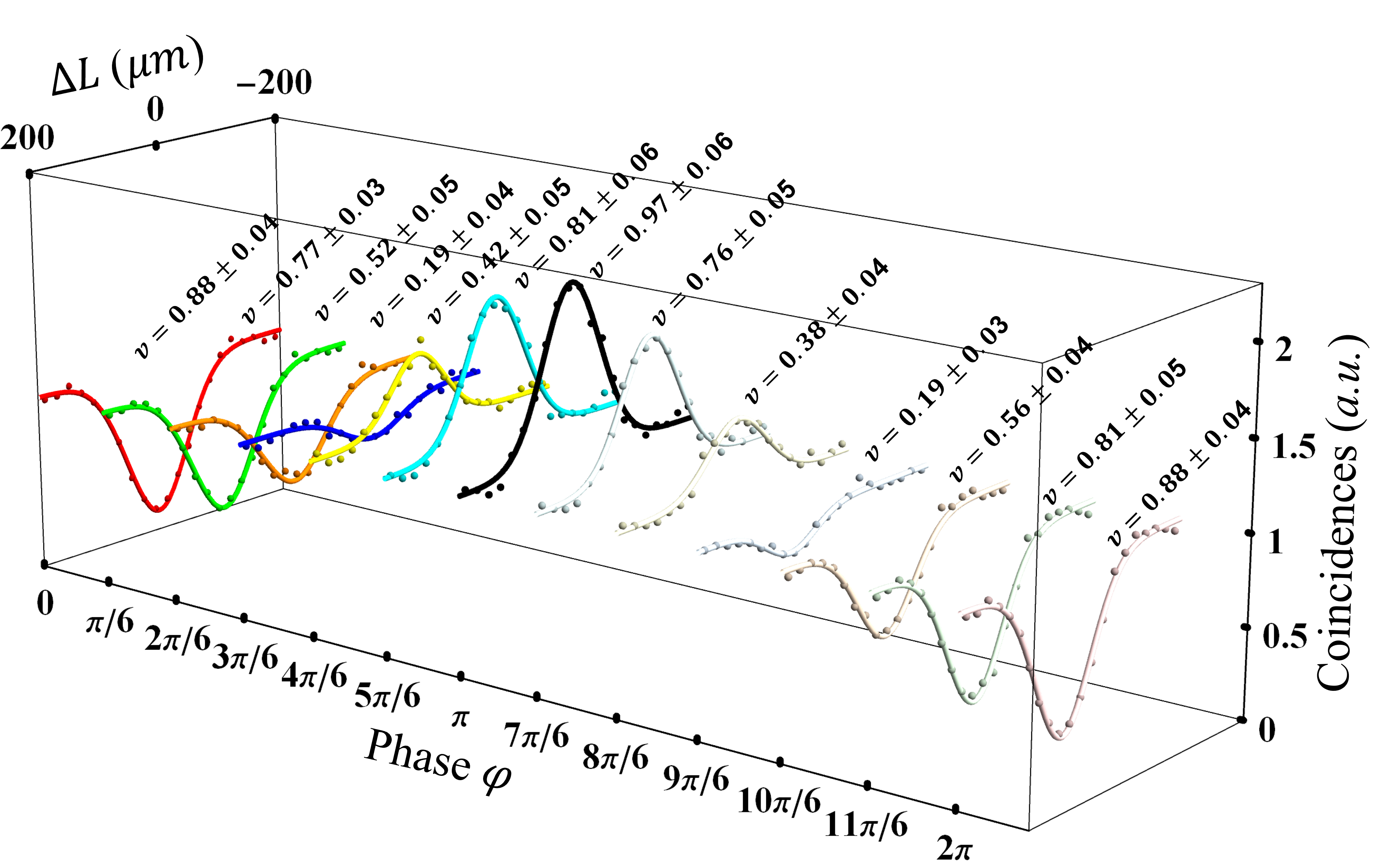}
\centering
\caption{\textbf{Two-fold normalized coincidence counts for different phase jumps.} The different colored data points and Gaussian fits correspond to phase ($\varphi$) changes every $\pi /6$ from 0 to $2 \pi$. 
${\Delta}L$ is the path length difference between the two photons before the BS.
The coincidences are normalized by dividing the coincidence rate at ${\Delta}L=0$ by that when ${\Delta}L$ is outside the HOM dip/peak.
Error bars are smaller than the point markers and therefore not visible in the plot.
}
\label{fig:visibility}
\end{figure*}

The scheme we propose is based on a Type-II SPDC source of photon pairs. Exploiting the fact that idler and signal photons are orthogonally polarized, we can spatially separate the two photons with a polarizing beamsplitter (PBS). The vertically polarized photons are then converted to horizontally polarized by a half-wave-plate, therefore the resulting two-photon state (in the transverse momentum degree of freedom) can be written as $\ket{\Psi}=\frac{1}{\sqrt{2}}\int d^2\mathbf{k}(\ket{\mathbf{k}}_s\ket{\mathbf{-k}}_i+\ket{\mathbf{-k}}_s\ket{\mathbf{k}}_i).$ Note that we have assumed here a perfectly collimated pump in the thin-crystal limit. A phase object in the far-field of the crystal, placed in the idler path, implements the transformation $\ket{\mathbf{k}}_i \rightarrow e^{\phi(\mathbf{k})}\ket{\mathbf{k}}_i$. When post-selecting on correlated pairs of momentum values (i.e., $\mathbf{k_0}$ and $-\mathbf{k_0}$), we obtain the state:  
\begin{equation}
    \ket{\Psi_{\varphi}}=\frac{1}{\sqrt{2}}(\ket{\mathbf{k_0}}_s\ket{\mathbf{-k_0}}_i+e^{i\varphi(\mathbf{k_0}) }\ket{\mathbf{-k_0}}_s\ket{\mathbf{k_0}}_i),
    \label{state}
\end{equation}
where $\varphi(\mathbf{k_0})=\phi(\mathbf{k_0})-\phi(-\mathbf{k_0})$, and we ignore a global phase factor.\newline

The state symmetry can be analyzed through Hong-Ou-Mandel (HOM) interference. When the two photons are incident on a beamsplitter (BS), with the important requirement that the number of reflections up to the exit port of the BS has the same parity for both photons, they exit from the same output port if the relative phase is $\varphi=0$ (bunching), while they always exit from different ports if $\varphi=\pi$ (anti-bunching). This can be immediately seen when measuring the coincidence counts between the two output paths $a$ and $b$ of the BS. More generally, for arbitrary phases $\varphi$, the (normalized) coincidence count rate is 
\begin{equation}
C(\varphi)=1-\cos(\varphi),
\label{eq:countrate}
\end{equation}
which can be inverted to obtain $\varphi$ (modulo $\pi$) that characterizes the state completely.

\section*{Results}
\begin{figure}[tbph!]
\includegraphics [width=0.8\linewidth]{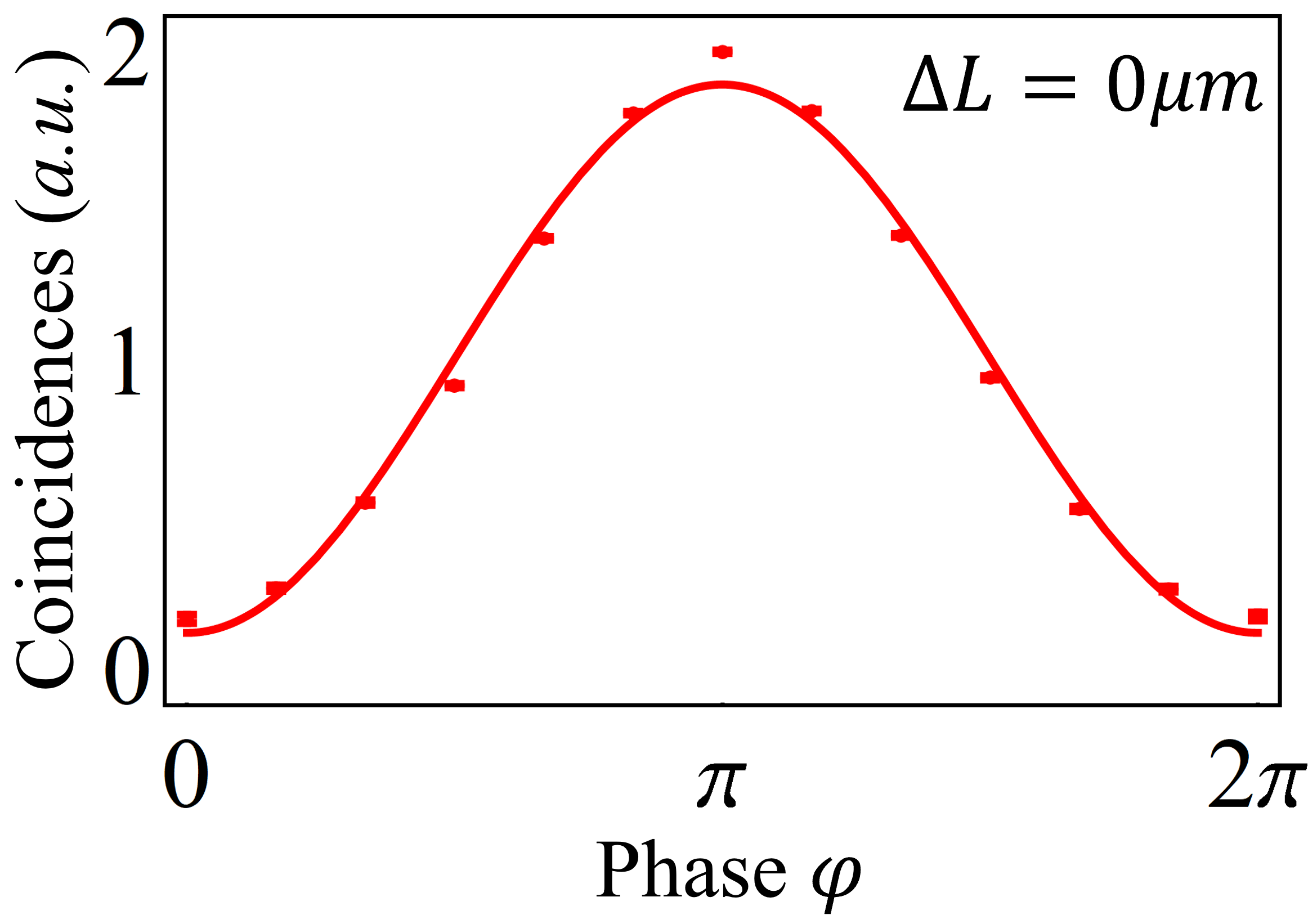}
\centering
\caption{\textbf{Coincidence counts as a function $\varphi$.} 
Plot showing the variation of coincidence counts as a function of the applied phase $\varphi$ when the path length difference ${\Delta}L$ between the two photons is 0. The equation of the fitted curve is $C(\varphi)=\alpha\left(1-\cos(\varphi)\right)+\beta$, where $\alpha=0.89\pm0.02$ and $\beta=0.12\pm0.03$.
The coincidences are normalized by dividing the coincidence rate at ${\Delta}L=0$ by that when ${\Delta}L$ is outside the HOM dip/peak.
Error bars are not visible, being smaller than the size of the data point circle.
}
\label{fig:coincidence}
\end{figure}

To experimentally test our scheme, we generated momentum entangled photon pairs in SPDC using a 5mm thick type-II PPKTP crystal pumped by a 405\,nm continuous wave laser (the detailed experimental setup is illustrated in Fig. \ref{fig:setup}). The frequency-degenerate photon pairs are split in idler and signal path by a polarizing beamsplitter (PBS). A reconfigurable liquid crystal spatial light modulator (SLM) allows us to apply different phase patterns on the idler path. In particular, we implemented phase jumps along a vertical line centered on the SPDC cone. Idler and signal photons are then made to interfere at a 50:50 BS (after the signal polarization has been rotated to horizontal). The number of mirrors in the two paths was chosen in order to keep momentum anti-correlation at the BS output ports, i.e., ensuring that both photons are subject to the same number of reflections. The two output modes (paths $a$ and $b$) were then coupled to single-mode fibers in such a way that, on each path, opposite values of the transverse momentum were selected. 

One thing to note is that this scheme requires photons with momentum $\mathbf{k}$ and $\mathbf{-k}$ to be present in both arms of the interferometer. Therefore, one cannot use a knife-edge prism or a D-shaped mirror to split the photons into two paths. This makes momentum entanglement manipulation using Type-I or Type-0 SPDC, where the photon pairs have the same polarization, much more difficult. The same technique can also be used for manipulating the transverse position entanglement between SPDC photons. This will require the parity of the number of reflections to be unequal in order to convert from a symmetric to an anti-symmetric position entangled state.

We recorded coincidence counts as a function of the path length difference $\Delta L$ between the two paths in the interferometer and for different phase jumps. The results are shown in Fig.~\ref{fig:visibility}, illustrating how applying a phase jump allows one to switch from two-photon bunching to two-photon anti-bunching, a clear indicator that the momentum entanglement has been converted from symmetric to anti-symmetric. The visibilities $v$ of the HOM dip in photon bunching and of the coalescence peak are defined as $v_\text{peak}=(C_{max}-C)/C$ and $v_\text{dip}=(C-C_{min})/C$, where $C_{max}$ and $C_{min}$ are the maximum and minimum coincidence counts at the peak and dip, respectively. $C$ is the coincidence count outside the dip/peak where the difference in the two path lengths is much larger than the coherence length of the SPDC photons.


In Fig.~\ref{fig:coincidence}, we verify that the coincidences at $\Delta L=0$ follow Eq.~\eqref{eq:countrate}. The visibilities for $\varphi=0$ and $\pi$ give a direct estimate of the fidelity between the realized state and the expected symmetric/antisymmetric entangled state. We obtained $v\sim88\%$ and $v\sim97\%$, for $\varphi=0$ and $\pi$, respectively. Intermediate cases correspond to the creation of entangled states which mimic two-particle states obeying anyonic statistics~\cite{sansoni2012two}.

{The visibilities do not quite reach 1 at $\varphi=0$ and $\pi$, which is due to the imperfections in the alignment and the BS not being exactly 50:50. We have found the alignment which gave the highest visibility in the peak often deviated slightly from the alignment which gave the highest visibility in the HOM dip; this can be a result of small imperfections in the alignment and SLM calibration. We have aligned the setup by maximizing the HOM peak, thus resulting in the visibility at $\varphi=\pi$ being higher than $\varphi=0$. Non-uniformity in the SLM will have only a small affect in our experiment as we are collecting photons from two small regions on the SLM. If photons were collected from larger regions by using multi-mode fibers or camera, then the SLM uniformity would need to be considered.}

\section*{Conclusions and outlook}
{Measurement of state symmetry using HOM interference has been demonstrated in various degrees of freedom, such as polarization \cite{PhysRevLett.76.4656}, frequency \cite{PhysRevLett.103.253601, fedrizzi2009anti, kaneda2019direct}, and Orbital Angular Momentum \cite{zhang2016engineering}.}
Here, we introduced and demonstrated a new method that allows, for the first time, the manipulation of transverse momentum entanglement between SPDC photon pairs through the use of a reconfigurable phase object (SLM), which allows to locally tune the state symmetry.
We also demonstrated how a HOM interferometer must be constructed in order to verify the momentum entanglement symmetry. Simultaneously generating a state with spatially variable symmetry with thousands of momentum entangled modes can be performed in the near future and directly observed with recently developed time-tagging camera technologies~\cite{Perenzoni2016,Nomerotski2019}.  This ability to manipulate and measure thousands of entangled modes would be greatly beneficial in quantum imaging protocols as well as in high-dimensional quantum communications, quantum information processing, and quantum simulations. However, this will require a high visibility multimode HOM interference in the spatial domain, a task that necessitates a careful engineering of SPDC spatial correlations, which requires the generation of a SPDC state with very high spatial correlation and compensation for the different SPDC cone sizes associated to the orthogonal polarizations.

\section*{Funding}
This work was supported by the Canada Research Chairs (CRC), the High Throughput and Secure Networks (HTSN) Challenge Program at the National Research Council of Canada, Canada First Research Excellence Fund (CFREF) Program, and Joint Centre for Extreme Photonics (JCEP). 

\section*{Acknowledgments}
The authors would like to thank Dilip Paneru and Fr\'ed\'eric Bouchard for valuable discussions.

\section*{Disclosures}
The authors declare no conflicts of interest.

\section*{Data Availability}
Data underlying the results presented in this paper are not publicly available at this time but may be obtained from the authors upon reasonable request.



\bibliography{refs}

\end{document}